\documentstyle[aps,prb]{revtex}
\begin{document}
\draft
\flushbottom

\twocolumn[
\title{Thermal conductivity across a twin boundary 
in $d$-wave superconductor}
\author{M. E. Zhitomirsky 
and M. B. Walker}
\address{Department of Physics, University of Toronto, Toronto, 
Canada M5S 1A7}
\date{\today}
\maketitle
\widetext \advance\leftskip by 57pt \advance\rightskip by 57pt
\begin{abstract}
We consider excitation spectrum near a twin boundary in an 
orthorhombic $d+s$ superconductor. The low-energy spectrum is highly 
sensitive to the presence of the small amount of 
$s$-wave component. Robustness 
of the bound states at the Fermi level with respect to impurities and 
an extra boundary potential is investigated. The role of Andreev 
transmission process for the low-temperature thermal conductivity 
across twin boundaries is studied for an impure superconductors. 
At very low temperatures the bulk part and the twin
boundaries part of the thermal conductivity have similar linear-$T$
dependences, whereas at intermediate temperatures the two contributions
behave like $T^3$ and $T^2$, respectively.

\end{abstract}
]\narrowtext

\section{Introduction}

The bound and extended quasiparticle states at twin boundaries in an 
orthorhombically twinned $d$-wave superconductor (like, supposedly, 
YBa$_2$Cu$_3$O$_{7-x}$) have recently been studied for the clean case. 
\cite{ZhW} An areal density of low energy twin-boundary bound states 
proportional to the magnitude of the $s$-wave component of the 
superconducting gap (which is a linear combination of tetragonal 
$s$-wave and $d$-wave components \cite{SR,Walker96}) was found, and a 
process of Andreev transmission (particle-hole conversion in 
transmission across the twin boundary) was shown to be the essential 
mechanism of transporting heat across the twin boundaries in the limit 
of sufficiently low temperatures. 

The present article considers the effect of disorder on the 
twin-boundary bound and extended states. A number of studies have shown 
the presence of impurities gives rise to a significant bulk density of 
zero energy excitations in the $d$-wave superconductor and that these 
have a significant effect on the bulk thermal conductivity.
\cite{Hotta,HG,Lee,Carb,PKM,Sun,Graf,HP} This paper 
shows that the enhanced density of states at the Fermi level affects 
also the transport of heat across the twin boundaries. Furthermore, in 
addition to giving a finite lifetime to the extended bulk excitations, 
impurities transform the twin-boundary bound states into decaying 
resonances. 

Theoretical calculations (see, e.g., Ref.~\onlinecite{Carb}) show that 
for a $d$-wave superconductor the low temperature $a$--$b$ plane 
penetration depth varies linearly with $T$ in the clean limit, but that 
impurity scattering can change this variation to $T^2$. Since this 
result depends essentially on the fact that the gap displays lines of 
nodes, and not on the $d$-wave symmetry, it is applicable to 
YBa$_2$Cu$_3$O$_{7-x}$ (YBCO), where the orthorhombic $A_{1g}$ gap is 
considered to be a mixture of $s$-wave and $d$-wave components. In the 
Born limit, the amount of scattering required to produce the $T^2$ 
variations of the penetration depth is so large that $T_c$ will be 
dramatically reduced, but in the unitary limit impurity scattering can 
give the $T^2$ temperature dependence without greatly affecting $T_c$. 
Very clean superconductors do show \cite{lambda} the low temperature 
$a$--$b$ plane penetration depth varying linearly with $T$ whereas 
detailed analysis \cite{Annet,Bonn} of the results from a number of 
other samples reveals a $T^2$-behavior. For these reasons, much recent 
work has analyzed the effect of the unitary-limit impurity scattering 
on the properties of high-temperature superconductors. 

Since a somewhat intuitive approach will be used below for 
calculation the thermal conductivity of the twin boundaries (e.g., in 
comparison with the Ambegaokar-Griffin \cite{Grif} approach to the 
calculations of bulk thermal conductivity), we review here the relevant 
qualitative ideas and their application to the description of other 
properties of high-temperature superconductivity. For a clean $d$-wave 
superconductor the density of states for the excitations in the 
superconducting state varies linearly with excitation energy at low 
energies and can be written in the form 
\begin{equation}
N_s(E) / N_0 \simeq E/\Delta_0 \ ,
\label{pureN}
\end{equation}
where $N_0$ is the normal state density of states and $\Delta_0$ is the 
maximum gap. In the strong scattering limit, the energy of low-lying 
excitations with wave-vector $\bf k$ acquires an imaginary part, i.e., 
$E_{\bf k}$ becomes $E_{\bf k}+i\Gamma$, thus the excitation possessing 
a lifetime $\tau = 1/(2\Gamma)$ ($\hbar=1$). The quantity $\Gamma$ is 
given by the solution of the following equation 
(see, e.g., Ref.~\onlinecite{PKM}) 
\begin{equation}
\Gamma^2 = \frac{n_i \Delta_0}{2N_0\ln(4\Delta_0/\Gamma)} \ ,
\end{equation}
where $n_i$ is the impurity concentration. Note that this result for 
the damping of an excitation is independent of the wave-vector and 
energy, and is valid for quasiparticles with energies less than $\Gamma$. 
Since, by the 
uncertainty principle, the low-energy excitations will have a spread in 
energy of the order of $\Gamma$, excitations with $E_{\bf k}$ from zero 
up to $\Gamma$ should contribute to the density of states at zero 
energy. Thus, taking account of Eq.~(\ref{pureN}), we have 
\begin{equation}
N_s(E) /N_0 \simeq\Gamma/\Delta_0 
\end{equation}
for the density of superconducting excitations at zero energy in the 
presence of unitary-limit impurity scattering. The exact result 
\cite{PKM} differs from this by a logarithmic factor which varies 
weakly with impurity concentration and which is often neglected in 
qualitative discussions.\cite{Lee} The low temperature specific heat 
associated with this constant density of states at low energies is 
($k_B=1$)
\begin{equation}
C_V = \frac{2\pi^2}{3} N_0 \frac{\Gamma}{\Delta_0} T \ .
\label{C}
\end{equation}
These ideas allow an estimate 
of the bulk thermal conductivity to be made as \cite{Graf}  
\begin{equation}
\kappa\! \sim\! C_V v_Fl\! \sim\! \left(N_0 \frac{\Gamma}{\Delta_0} T\right) 
v_F\left(v_F\frac{1}{\Gamma}\right)  
 = \frac{\pi^2}{3}\frac{v_F^2N_0}{\Delta_0} T \ .
\end{equation}
Note that, like the result for the electrical conductivity in the 
unitary limit of impurity scattering,\cite{Lee} this is a universal 
result. The parameter $\Gamma$, which depends on $n_i$, cancels out and 
the thermal conductivity is independent of the impurity concentration 
and the strength of impurity potential. Recently, the universal heat 
transport has been observed \cite{Tail} in untwinned crystals 
YBa$_2$(Cu$_{1-x}$Zn$_x$)$_3$O$_{6.9}$. Quantitative comparison of 
these data with the theory suggests that quasiparticle scattering on Zn 
impurities in YBCO is close, but does not coincide completely with the 
unitary limit. 

At higher energies $\Gamma\ll E\ll\Delta_0$ the density of states 
reverts to it's clean-limit value (\ref{pureN}), whereas the 
relaxation rate of superconducting quasiparticles acquires energy 
dependence $\Gamma_S=\Gamma_N\Delta_0/E$ (Ref.~\onlinecite{Lee}),
where the normal 
state relaxation rate in the unitary limit is $\Gamma_N=n_i/\pi N_0$. 
In the corresponding temperature range $\Gamma\ll T\ll \Delta_0$ the 
electronic thermal conductivity can be calculated by an elementary 
Boltzman equation approach\cite{Bardeen} 
\begin{equation}
\kappa = \sum_{\bf k}\frac{(E_{\bf k}v_{\bf k}\cos\theta_{\bf k})^2}
{T\Gamma_S}
\left(-\frac{\partial f_{\bf k}^0}{\partial E_{\bf k}}\right) .
\end{equation}
provided 
the correct energy dependence of 
$\Gamma_S$ given above is taken into account. Using the fact that
near the gap nodes the quasiparticle group velocity reduces
to $v_F$, we find after integration
\begin{equation}
\kappa = \frac{7\pi^4}{15}\frac{v_F^2N_0}{\Gamma_N\Delta_0^2} T^3 
\ .
\end{equation}
Thus, above the universal linear-$T$ behavior at very low temperatures
the heat current in $d$-wave superconductor
varies as $T^3$, which is consistent with the numerical
data presented in Ref.~\onlinecite{Graf}.

\section{Quasiparticle states near twin boundary}

In a tight-binding model on a square lattice (supposedly reproducing in 
a more or less correct fashion a single CuO$_2$ plane in YBCO),  the 
Green function describing the excitations in the superconducting state 
is determined by the equation 
\begin{equation}
\sum_{k} \left[\hat{1}\tilde{z} \delta_{ik} - 
\hat{\epsilon}_{ik}\right] 
\hat{G}(k,j,z) = \hat{1}\delta_{ij} \ .
\label{Green}
\end{equation}
Singlet state pairing is assumed and a ``hat'' indicates a $2\times 2$ 
matrix in particle-hole space. Disorder is included in this model by 
considering impurities on randomly chosen lattice sites, each impurity 
being described by an additional potential $u$. Impurity scattering is 
treated in the self-consistent $t$-matrix approximation. As a result, 
the quantity $\tilde{z}$ must be determined from the equation 
\begin{equation}
\tilde{z} = z - \frac{n_iu^2{\cal G}_0(\tilde{z})} 
{1-u^2{\cal G}_0(\tilde{z})^2} \ ,
\label{z}
\end{equation}
where
\begin{equation}
{\cal G}_0(\tilde{z}) = \case{1}{2}\text{Tr}\hat{G}(i,i,z) \ .
\label{tr}
\end{equation}
The right hand side of Eq.~(\ref{tr}) is assumed to be independent of 
site index $i$, $z$ is the usual Matsubara frequency, and the $s$-wave 
component of the gap is assumed negligible in comparison with the 
$d$-wave component in arriving at Eq.~(\ref{z}). 

In the unitary limit ($u\rightarrow\infty$) detailed study of 
Eq.~(\ref{z}) shows that for low energy excitations, i.e., excitations 
having an energy less than $\Gamma$, the quantity $\tilde{z}$ is given 
by 
\begin{equation}
\tilde{z} = z \pm i\Gamma
\label{z1}
\end{equation}
where $\Gamma$ has been defined in the Introduction, and the upper and 
the lower signs refer to the upper and lower half plane, respectively. 

The matrix $\hat{\epsilon}_{ik}$ contains the same quantities that were 
used in our previous model of a clean superconductor containing a twin 
boundary:\cite{ZhW} 
\begin{equation}
\hat{\epsilon}_{ik} = \hat{\tau}_3\left(-t_{ik} - \mu\delta_{ik} + 
U_0\delta_{ik}\delta_{i\text{B}}\right) + \hat{\tau}_1 \Delta_{ik}
\end{equation}
where $\delta_{i\text{B}}$ is unity for $i$ on the twin boundary and 
zero otherwise and $\Delta_{ik}$ is assumed real.

When the site indices are incorporated into the matrix notation, 
Eq.~(\ref{Green}) becomes 
\begin{equation}
\left[\hat{1}\tilde{z} - \hat{\epsilon}\right] \hat{G} = \hat{1} \ .
\end{equation}
where $\tilde{z}$ is given by Eq.~(\ref{z1}). Since $\hat{\epsilon}$ is 
a real symmetric matrix, it can be diagonalized by a unitary 
transformation, say $S$, giving 
\begin{equation}
\hat{G} = S^{\dag} (\hat{1}\tilde{z} - \hat{\Lambda})^{-1} S \ ,
\end{equation}
where $\hat{\Lambda}$ is the diagonal matrix of the eigenvalues of 
$\hat{\epsilon}$. The excitation energies are found from the poles of 
$\hat{G}$ when $z\rightarrow E+i0$. Thus, in the low-energy limit, if 
$\Lambda_i$ is an eigenvalue of $\epsilon$, there will be an excitation 
with energy 
\begin{equation}
E = \Lambda_i + i\Gamma \ .
\end{equation}
Therefore, all low energy excitations, including those which would be 
bound to the surface in the absence of impurity scattering, will have 
the same lifetime $\tau =1/2\Gamma$.

\subsection*{Diagonalization of the tight-binding Hamiltonian}

The diagonalization of $\hat{\epsilon}$ is essentially the same as in 
Ref.~\onlinecite{ZhW}. Here we give some details not previously 
explicitly described, and in particular calculate transmission 
coefficients across the twin boundary in the superconducting state. 

We do not solve the problem for the superconducting order parameter 
near the twin boundary self-consistently, but instead substitute a 
``guess'' order parameter into the Bogoliubov-de Gennes (BdG) equations 
\begin{equation}
E \psi_i = \hat{\epsilon}_{ik} \psi_k
\label{BdG}
\end{equation}
to study the qualitative features of the excitation spectrum. 
Definitions of nearest-neighbor hopping and pairing amplitudes are 
given in Fig.~1. In accordance with the orthorhombic symmetry of 
CuO$_2$ planes in YBCO these amplitudes are different along $a$ and $b$ 
axes: $t_{1,2}=t(1\pm\epsilon)$ and $\Delta_{1,2}=\Delta(1\pm\delta)$. 

In the bulk of each twin, quasiparticles are plane waves with excitation 
energy
\begin{equation}
E_{\bf k} = \sqrt{\varepsilon({\bf k})^2 + \Delta({\bf k})^2} \ ,
\end{equation}
where the normal state quasiparticle dispersion in the right (left) 
twin is given by 
\begin{equation}
\varepsilon({\bf k}) = 
-4t\cos\frac{k_xa}{\sqrt{2}}\cos\frac{k_ya}{\sqrt{2}} \mp 
 4t\epsilon\sin\frac{k_xa}{\sqrt{2}}\sin\frac{k_ya}{\sqrt{2}}-\mu \ ,
\label{dispersion}
\end{equation}
and the mixed symmetry superconducting gap is 
\begin{equation}
\Delta({\bf k}) = 2\Delta 
\sin\frac{k_xa}{\sqrt{2}}\sin\frac{k_ya}{\sqrt{2}}
\mp 2\Delta\delta 
\cos\frac{k_xa}{\sqrt{2}}\cos\frac{k_ya}{\sqrt{2}} 
 \ , \label{gap}
\end{equation}
where the first term corresponds to a $d$-wave component, while the 
second term describes an extended $s$-wave component (the only $s$-wave 
harmonic in our model) which admixes with different signs in the two 
crystal twins. At half-filling the maximum amplitudes of the two 
harmonics are $\Delta_d=\Delta$ and $\Delta_s=\Delta\delta < \Delta_d$. 

Treating quasiparticle scattering in both the normal and the 
superconducting states one can employ the translational invariance 
along the twin boundary, which leads to conservation of the parallel 
component of the momentum. For low-energy elastic processes we need to 
determine change in the normal component of the momentum of a scattered 
excitation. From (\ref{dispersion}), the normal components of the Fermi 
momenta for incoming and outgoing electrons with a given parallel 
component $k_y$ are $k_{\pm x}^>=\pm k+q$, in the right twin, and 
$k_{\pm x}^<=\pm k-q$, in the left twin (see Fig.~2). Parameters $k$ 
and $q$ are defined by 
\begin{equation}
\tan\frac{qa}{\sqrt{2}}=\epsilon\tan\frac{k_ya}{\sqrt{2}}\ ,  \ \ \ \ 
\cos\frac{ka}{\sqrt{2}}=\frac{-\mu\cos(qa/\sqrt{2})}
{4t\cos(k_ya/\sqrt{2})}\ .
\end{equation}
The normal components of the Fermi velocities for all these states 
have the same absolute value $v_{Fx}\! =\! 
2\sqrt{2}ta\sin(ka/\sqrt{2})\cos(k_ya/\sqrt{2})/\cos(qa/\sqrt{2})$. 

The two-component BdG eigenfunctions on the left and right sides, 
$\psi^<_i$ ($x<0$) and $\psi^>_i$ ($x>0$) in the twin boundary problem, 
 are still given by a combinations of plane waves, therefore among BdG 
equations (\ref{BdG}) a special care is required only for those with 
$i$ on the twin boundary. Subtracting from the difference equation 
(\ref{BdG}) for $i\in\text{TB}$ a uniform part and making Fourier 
transform over $y$ we obtain two conditions: (i) continuity equation 
$\psi^<_0=\psi^>_0$ and (ii) an additional equation ``on the 
derivative,'' which in the leading order in in superconducting gap, 
i.e., neglecting corrections $O(\Delta^2/\varepsilon_F)$ to the energy, 
has the following form 
\begin{equation}
\psi^>_1e^{-iqa/\sqrt{2}}-\psi^<_1e^{iqa/\sqrt{2}} 
= \frac{U_0\cos(qa/\sqrt{2})}{2t\cos(k_ya/\sqrt{2})}\,\psi_0 \ .
\label{discrete}
\end{equation}
In the following analysis we will also use the symmetry of the twin 
boundary which is expressed by the relations 
$t_{\hat{\sigma}i\hat{\sigma}j}=t_{ij}$ and 
$\Delta_{\hat{\sigma}i\hat{\sigma}j}=-\Delta_{ij}$, where 
$\hat{\sigma}$ is reflection in the twinning plane (not a Pauli 
matrix). The BdG equations (\ref{BdG}) are invariant under the combined 
transformation $\hat{U}=\tau_3\hat{\sigma}$, $\tau_3$ being the Pauli 
matrix which acts in the particle-hole space. All solutions are, hence, 
classified by their parity with respect to $\hat{U}$. 

In the normal state, quasiparticles move freely across the twin 
boundary for $U_0=0$ and acquire a finite scattering amplitude 
$r=\alpha/(i-\alpha)$ for $U_0>0$. 

Let us consider a scattering process below $T_c$ during which a 
quasiparticle approaching the twin boundary from the right with the 
wave-vector ${\bf k}_-^>$ is reflected into an out-going state with the 
wave-vector ${\bf k}_+^>$ on the same side and is transmitted into 
out-going states with ${\bf k}_-^<$ and ${\bf k}_+^<$ on the left side 
(Fig.~2). We define $\Delta_-=\Delta({\bf k}_-^>)$, 
$\Delta_+=\Delta({\bf k}_+^>)$, $\Delta({\bf k}_-^<)=-\Delta_+$, and 
$\Delta({\bf k}_+^<)=-\Delta_-$, the last two relations following from 
the symmetry of the superconducting state on the twin 
boundary.\cite{Walker96} 

There are three cases for the quasiparticle energy $E$ to be 
considered, (i) $E<\Delta_{\text{min}}$, (ii)  
$\Delta_{\text{min}}<E<\Delta_{\text{max}}$, and (iii) 
$\Delta_{\text{max}}<E$, where $\Delta_{\text{min}} = 
\text{min}(|\Delta_-|,|\Delta_+|)$ and 
$\Delta_{\text{max}}=\text{max}(|\Delta_-|,|\Delta_+|)$.

\subsubsection{Bound states, $E<\Delta_{\text{min}}$ }

The bulk states in this energy range are absent and the only 
possibility is states localized in the vicinity of the twin boundary. 
The wave function of a bound state in the right twin is given by 
\begin{eqnarray}
\psi^>_i\!&=&\!\!\left(\!\!\begin{array}{c}\Delta_-\\E+i\Omega_-
\end{array}\!\!\right)\! e^{i{\bf k}^>_-\cdot{\bf r}_i-\kappa_-x}\!+\! 
R\left(\!\!\begin{array}{c} \Delta_+\\E-i\Omega_+\end{array}\!\!\right) 
 \!e^{i{\bf k}^>_+\cdot{\bf r}_i-\kappa_+x}     
\nonumber \\ & & \Omega_\pm=\sqrt{\Delta_\pm^2-E^2} \ , \ \ \ 
\kappa_\pm = \Omega_\pm/v_{Fx} 
\label{bs}
\end{eqnarray}
whereas on the left side $\psi^<_i$ is obtained from $\psi^<_i = 
\pm\tau_3\psi^>_{\hat{\sigma}i}$ for even (odd) eigenstates of the 
operator $\hat{U}$. The ratio of the particle-hole components of the 
wave-function is determined from homogeneous BdG equations, while 
energy $E$ and ``reflection'' coefficient $R$ have to be found from the 
continuity condition and Eq.~(\ref{discrete}). Let us consider these 
equations in more detail for the odd symmetry solutions. From the 
continuity of the wave function one finds 

\begin{equation}
\Delta_- + R \Delta_+ = 0 \ ,
\label{eb1}
\end{equation}
whereas Eq.~(\ref{discrete}) yields
\begin{equation}
E + i\Omega_- - R(E-i\Omega_+) = i\alpha [E+i\Omega_- + R(E-i\Omega_+)] 
\ ,
\label{eb2}
\end{equation}
where $\alpha= U_0 a /\sqrt{2}v_{Fx}$. From the real part of the last 
equation we find for the bound state energy
\begin{equation}
E = \pm \frac{2\alpha|\Delta_-|}
{\sqrt{4\alpha^2+[1-R-\alpha^2(1+R)]^2}} \ .
\label{eb3}
\end{equation}
The sign in this equation should be found by substituting expression
(\ref{eb3}) back 
into Eq.~(\ref{eb2}). This gives 
\begin{equation}
|1-R-\alpha^2(1+R)| = \mp [1-R-\alpha^2(1+R)]
\label{eb4}
\end{equation}
and 
\begin{equation}
\frac{R}{|R|} = \pm \frac{[\alpha^2(1+R)+1-R]}{|\alpha^2(1+R)+1-R|} \ . 
\label{eb5}
\end{equation}
As follows from Eqs.~(\ref{eb4}) and (\ref{eb5}) the two cases $|R|<1$ 
and $|R|>1$ transform into each other under $R\leftrightarrow1/R$. 
Consequently, we have to consider only $-1<R<1$. 

For $-1<R<0$ or $\Delta_+\Delta_->0$ we find from Eq.~(\ref{eb4}) the 
upper sign if $\alpha^2>(1-R)/(1+R)$ and the lower sign in the opposite 
case. On the other hand Eq.~(\ref{eb5}) gives always the lower sign. 
Therefore, the energies of the bound states of odd symmetry are given 
by (\ref{eb3}) with the minus sign as far as the boundary potential is 
weak and $\alpha$ is small. The dispersion of bound state energies 
disappears at $U_0=0$. The condition $\Delta_+\Delta_->0$ is satisfied 
in the vicinity of the gap nodes, e.g., for 
$k_{y\text{C}}<k_y<k_{y\text{A}}$ and for $k_y > k_{y\text{D}}$ and 
$k_y<k_{y\text{B}}$ in Fig.~2. The number of the bound states is, thus, 
proportional to the degree of the orthorhombicity of the 
superconducting gap or $\Delta_s/\Delta_d$, where $\Delta_s$ and 
$\Delta_d$ are amplitudes of $s$- and $d$-wave gaps.\cite{ZhW} 
Observation of the zero-energy peak in the local density of states near 
twin boundary would be a direct experimental evidence of the mixed 
symmetry gap in YBCO. 

The critical value of the boundary potential $U_0$, which destroys the 
bound states, is different for states propagating normally to the 
twinning plane and for states propagating nearly parallel. In the 
former case $\alpha\sim U_0/\tilde{\varepsilon}$, where 
$\tilde{\varepsilon}$ is geometrical average of the Fermi energy 
$\varepsilon_F$ and the band width. The parameter $\alpha$ is, 
therefore, quite small and cannot exceed the critical value 
$\alpha_c=1$. The bound states lie near the Fermi level having a weak 
dispersion given by Eq.~(\ref{eb3}). In the second case of the bound 
states with momenta nearly parallel to the twin boundary, perpendicular 
component of the Fermi velocity is very small and $\alpha$ is 
significantly enhanced. Bound states near ``vertical'' nodes in Fig.~2 
disappear completely for $U_0/\tilde{\varepsilon} > \Delta_s/\Delta_d$. 

For $0<R<1$ or $\Delta_+\Delta_-<0$, that is, in the ``out of nodes'' 
region, Eq.~(\ref{eb5}) yields the upper sign. This is compatible with 
Eq.~(\ref{eb4}) if $\alpha^2>(1-R)/(1+R)$. Arbitrary weak boundary 
potential can produce bound states for the part of the Fermi surface 
with $R\approx1$. Total number of such bound states is, however, 
proportional to $\alpha$. As a substitution of $R\approx1$ into 
(\ref{eb3}) shows, these bound states split from the bottom of the 
continuum of bulk quasiparticles, i.e., have energies slightly below 
$\Delta({\bf k})$, which does not vanish in this region of the Fermi 
surface.  Therefore, such a possibility being sensitive to the value of 
the boundary potential is  less important for the thermodynamic of the 
system. 

Completely analogous consideration of the even symmetry states predicts 
the upper sign in Eq.~(\ref{eb3}) and the same condition on the two 
gaps 
\begin{equation}
\Delta_+\Delta_- > 0 
\end{equation}
in order to have bound quasiparticle states on the twin boundary.

Disorder provides an additional mechanism for broadening the 
zero-energy peak in the local density of states on twin boundary. The 
critical impurity concentration, which smears out completely bound 
states, is obtained by the following arguments. The characteristic 
space extension of bound states is $\xi_s=v_F/\pi\Delta_s$. In the clean 
limit they can be considered as quasiparticles undergoing multiple 
Andreev reflections in the superconducting well on the twin boundary. 
In the presence of impurity scattering, the coherence length $\xi_s$ 
has to be compared to the mean free path $l=v_F\tau=v_F/2\Gamma$. For 
$\xi_s\ll l$, a quasiparticle is reflected many times before being 
wiped out from the twin boundary region by a scattering on an impurity. 
In this case, though being transformed into resonances, the quasibound 
states are still well defined. For $\xi_s\geq l$, a quasiparticle 
scatters to the continuum very quickly and the twin boundary resonances 
as well as the zero-energy peak disappear. Thus, it is possible to 
destroy the zero-energy peak in the local density of states by 
increasing number of impurities to the critical value given by 
$\Gamma_c\sim\Delta_s$.

\subsubsection{Extended states with $|\Delta_-|<E<|\Delta_+|$}

In this energy range, an excitation, particle or hole, can approach the 
twin boundary from the bulk, but it cannot be reflected back into the 
twin with the wave-vector ${\bf k}_+^>$, since $E<|\Delta_+|$. The only 
allowed processes are Andreev reflection\cite{Andreev} and Andreev 
transmission,\cite{ZhW} which involve particle-hole transformations. 
The wave function of such a quasiparticle is given for $x>0$  by 
$$
\psi^>\!=\!\!
\left(\!\!\begin{array}{c}\Delta_-\\E_-\end{array}\!\!\right)\! 
e^{i{\bf k}^>_-\cdot{\bf r}}\!+\! 
A\left(\!\!\begin{array}{c} E_-\\\Delta_-\end{array}\!\!\right) 
 \!e^{i{\bf k}^>_-\cdot{\bf r}} + B 
\left(\!\!\begin{array}{c} \Delta_+\\E_+\end{array}\!\!\right) 
 \!e^{i{\bf k}^>_+\cdot{\bf r}-\kappa_+x}     
$$
and for $x<0$
\begin{equation}
\psi^<\!=\!\!
A'\left(\!\!\begin{array}{c} -\Delta_+\\E_+\end{array}\!\!\right) 
 \!e^{i{\bf k}^<_-\cdot{\bf r}+\kappa_+x}
 + B' 
\left(\!\!\begin{array}{c} E_-\\ -\Delta_-\end{array}\!\!\right) 
 \!e^{i{\bf k}^<_+\cdot{\bf r}} \ .
\label{And}
\end{equation}
We have neglected in the above formulas small shifts of the wave 
vectors from the Fermi surface for propagating 
states and defined 
\begin{equation}
E_-=E-\sqrt{E^2-\Delta_-^2} \ , \ \ \ 
E_+=E-i\sqrt{\Delta_+^2-E^2} \ .
\label{E+}
\end{equation}
Eqs.~(\ref{And}) describe an incident electron-like quasiparticle with 
momentum ${\bf k}\approx{\bf k}_-^>$ approaching the twin boundary from 
the right, a reflected hole with wave vector ${\bf k}\approx{\bf 
k}_-^>$ and transmitted hole with ${\bf k}\approx{\bf k}_+^<$. There 
are also two damped waves with wave vectors ${\bf k}_+^>$ and ${\bf 
k}_-^<$. We also took into account the odd symmetry of the gap with 
respect to the reflections in the twinning plane. 

Parameters $A$, $A'$, $B$, and $B'$ have to be determined from the 
continuity of the wave function and Eq.~(\ref{discrete}). For zero 
boundary potential one finds $B=B'=0$ so that only exponentially 
decaying wave exists for $x<0$. Hence, the transmission coefficient 
through the twin boundary for quasiparticles with 
$|\Delta_-|<E<|\Delta_+|$ vanishes. 

For $U_0\ne0$ there is a nonzero probability of an incident electron at 
${\bf k}_-^>$ to be transformed into an outgoing hole of wave-vector 
${\bf k}_+^<$ on the opposite side of the boundary, which is a kind of 
Andreev transmission. Remarkably, the boundary barrier $U_0>0$, which 
causes reflection of the quasiparticles from the twin boundary in the 
normal state, allows transmission of the low-energy excitations in the 
superconducting state. The particle-hole transmission coefficient to 
the second order in a small parameter $\alpha$  is 
\begin{equation}
w_{\text{ph}} = |B'|^2 = w_0 \frac{\Delta_+^4(E_-^2-\Delta_-^2)^2}
{|\Delta_-\Delta_+ + E_-E_+|^4} \ ,
\label{wph}
\end{equation}
where $w_0 = 4\alpha^2 = 2(U_0a/v_F)^2$. For $E\ll\Delta_+$ the 
transmission coefficient simplifies to
\begin{equation}
w_{\text{ph}} \approx  w_0 (E^2-\Delta_-^2)/E^2 \ .
\label{wpha}
\end{equation}

\subsubsection{Extended states with $E>\Delta_{\text{max}}$}

The wave function of scattered quasiparticle on the left and right 
sides is given by the same expression (\ref{And}) except for the 
decaying factors $e^{\pm\kappa_+x}$, which are absent now, and $E_+$ 
defined as $E_+=E-(E^2-\Delta_+^2)^{1/2}$ instead of Eq.~(\ref{E+}). 
The outgoing quasiparticle flow on the left side consists now of two 
parts corresponding to electron and hole excitations. Therefore, the 
transmission coefficient has the particle-hole contribution 
$w_{\text{ph}}$ of Eq.~(\ref{wph})  and the particle-particle 
contribution 
\begin{equation}
w_{\text{pp}} = |A'|^2 \frac{\Delta_+^2-E_+^2}{\Delta_-^2-E_-^2}
=  \frac{(\Delta_+^2-E_+^2)(\Delta_-^2-E_-^2)}{(\Delta_-\Delta_+ + E_-E_+)^2} 
\ ,  
\end{equation}
where the second equality is obtained for $\alpha\ll1$. For high
energies $E\gg\Delta_{\text{max}}$ in this case the particle-particle
transmission coefficient $w_{\text{pp}}\approx 1$ and 
$w_{\text{ph}}=O(\alpha^2)$.

\section{Thermal resistance of twin boundary}

In the case of dirty superconductor with 
strong impurity scatters, the width $\Gamma$ of the 
low-lying energy levels has an important implication on the thermal
conductivity through the twin boundaries. For $\Gamma>\Delta_s$, 
excitations with energies approximately equal to $\Gamma$ will be 
the most important carriers of heat at low temperatures, completely
analogous to the case of the bulk thermal conductivity (see Introduction).
For these excitations $w\approx1$ (if $\alpha\ll1$)
and the flow of the heat will 
not be much affected by the presence of the twin boundaries even in
the low-temperature limit.

For $\Gamma<\Delta_s$, the situation is different. At low temperatures
($T\ll\Gamma$) the heat will again be carried primarily by excitations
with energies $E\approx\Gamma$ and the mechanism of the Andreev 
transmission discussed above produces a nonzero flow of heat
across the twin boundary. Following Andreev, \cite{Andreev}
the heat current across the boundary is given by the expression
\begin{equation}
W = \frac{2}{L^3} \sum_{\bf k}\mbox{}^{^\prime}
E_{\bf k} n_{\bf k} v_{kx} w_{\bf k} \ .
\label{W}
\end{equation}
Here, the prime attached to the sum over momenta means that only 
excitations with $x$ component of the group velocity satisfying 
$v_{kx}<0$ are summed over; the $x$-axis is the axis normal to the twin 
boundary. Also, excitations with wave vectors near ${\bf k}_{\text{B}}$ 
and ${\bf k}_{\text{D}}$ in Fig.~2 are neglected since their group 
velocities toward the twin boundary are small. The factor $2$ accounts 
for the contribution of two spin directions.

For excitations near ${\bf k}_{\text{A}}$ and ${\bf k}_{\text{C}}$ 
(Fig.~2), the normal projection of the group velocity is 
\begin{equation}
v_{kx} = v_F \sqrt{E_{\bf k}^2-\Delta_{\bf k}^2}/ E_{\bf k} \ .
\end{equation}
Thus, using Eq.~(\ref{wpha}) for the transmission coefficient we
have
\begin{equation}
v_{kx} w_{\bf k} = v_F w_0 \left(
\frac{\sqrt{E^2_{\bf k}-\Delta^2_{\bf k}}}{E_{\bf k}} \right)^3 \ .
\label{app}
\end{equation}

In integrating over $\bf k$ [on the right hand side of Eq.~(\ref{W})]
in the neighborhood of ${\bf k}_{\text{A}}$
we approximate
\begin{equation}
E^2_{\bf k} = (v_F \delta k_x)^2 + (v_\Delta \delta k_y)^2 \ ,
\ \ \ \  \Delta_{\bf k} = v_\Delta \delta k_y
\end{equation}
and $|v_\Delta|\sim \Delta_0/k_F\ll v_F$. It is easily seen that the 
effect of the factor in brackets in Eq.~(\ref{app}), when averaged over 
$k$ as in Eq.~(\ref{W}), is to give a factor of the order of unity. For 
example, if the exponent of the factor in square brackets were 2 rather 
than 3, evaluation of the sum over $k$ in Eq.~(\ref{W}) shows that the 
correct result could be obtained by the substitution $v_{kx}w_{\bf 
k}=\case{1}{2}v_Fw_0$ for Eq.(\ref{app}). Since the exponent is 3 not 
2, we make a replacement 
\begin{equation}
v_{kx}w_{\bf k}= fv_Fw_0 \ ,
\end{equation}
where $f$ will be close to, but somewhat less than unity. Impurity 
scattering could not change this factor by very much.
With this substitution and extending the sum over $k$ in Eq.~(\ref{W})
to all $k$, we find
\begin{equation}
W=\case{1}{4}fv_Fw_0\left(\frac{2}{L^3}\sum_{\bf k}E_{\bf k}n_{\bf k}\right).
\label{W1}
\end{equation}
Note, that the quantity in the bracket is $\bar{E}-\mu N$, where
$\bar{E}$ is the mean energy of the system.

If the temperature on both sides of the twin boundary is the same,
the heat current $W$ is balanced by a heat current of the same
magnitude in the opposite direction. Thus, in the presence of
a temperature difference across the twin boundary, there will be a net
heat current given by
\begin{equation}
Q=\frac{\partial W}{\partial T}\delta T=
\kappa_{\text{TB}}\frac{\delta T}{d}\ ,
\end{equation}
where $d$ is an average spacing between twin boundaries. Hence, the 
thermal conductivity, if limited by twin boundary resistance, will 
have the low-temperature form
\begin{equation}
\kappa_{\text{TB}} = \case{1}{2} C_V v_F \lambda_{\text{eff}} \ ,
\end{equation}
where
\begin{equation}
C_V 
= T \left(\frac{\partial S}{\partial T}\right)_{V,N}
\approx \frac{\partial }{\partial T} (\bar{E}-\mu N) 
\label{Cv}
\end{equation}
and an effective mean free path
\begin{equation}
\lambda_{\text{eff}} = \frac{f}{2} w_0 d
\label{d}
\end{equation}
has been defined.
In writing Eq.~(\ref{Cv}), a weak temperature 
dependence of the chemical potential has been neglected.

Note, that since $v_F$ and $\lambda_{\text{eff}}$ do not depend on $T$, 
the temperature dependence of the twin boundary thermal conductivity 
will be that of the specific heat.

In the unitary limit for impurity scattering, the low-temperature 
specific heat at
$T<\Gamma$, has the form (\ref{C}). 
Thus, the thermal conductivity due to the twin boundaries has the same 
linear-$T$ dependence at low temperatures as the bulk thermal 
conductivity $\kappa_{\text{bulk}}$. If we now consider a sample with 
one predominant orientation of twin boundaries, say parallel to (110), 
then the thermal resistance in the perpendicular direction is additive 
and for the combined thermal conductivity from boundaries and 
impurities we have 
\begin{equation} 
\kappa_{\perp}^{-1} = (\kappa_{\text{TB}})^{-1} + 
(\kappa_{\text{bulk}})^{-1}\ .
\end{equation} 
The heat flow in the parallel direction is unaffected by twin boundaries  
and $\kappa_{\parallel} =\kappa_{\text{bulk}}$. Such 
anisotropy of in-plane thermal conductivity disappears for samples 
with equal weights of two types of twins. 

For $\Gamma\ll T\ll\Delta_s$, the specific heat 
coincides with that of a clean $d$-wave superconductor
$C_V=9\zeta(3)N_0T^2/\Delta_0$. Thus both in this case and in the case 
$T>\Delta_s$, when the scattering can be treated in the Born 
approximation, the twin boundary thermal conductivity varies as $T^2$. 
For the sake of comparison we note that in the unitary scattering limit and 
$\Gamma\ll T\ll\Delta_0$ the bulk thermal conductivity varies as 
$T^3$ (see Introduction), whereas in the Born limit\cite{HP} 
and for $T\ll\Delta_0$, $\kappa_{\text{bulk}}$ varies as $T$.

\section{Conclusions}

In the previous work\cite{ZhW} on a clean $d+s$ orthorhombic 
superconductor, such as YBCO, we have demonstrated the existence of 
bound zero-energy excitations at the twin boundaries and identified an 
Andreev transmission process responsible for heat conduction across 
the twin boundaries at low temperatures. When strong impurity 
scatters are added, a new energy scale $\Gamma$ appears, which is the 
relaxation rate of the lowest energy quasiparticles. In this paper we 
show that the twin boundary bound states remain well-defined 
resonances contributing a zero-energy peak to the local density 
of states at the 
twin boundary provided $\Gamma<\Delta_s$, where $\Delta_s$ is an 
amplitude of the $s$-wave component of the gap. Furthermore, we 
calculate the transmission coefficient describing the transmission of 
excitations across the twin boundaries, and hence evaluate the thermal 
conductivity of twin boundaries. For $\Gamma>\Delta_s$, the twin 
boundaries present resistance for the flow of heat due to an extra
boundary potential only, i.e., the same as in the normal state. For 
$\Gamma<\Delta_s$, on the other hand, there are two cases to be considered. If 
$T\ll\Gamma$, the twin boundary thermal conductivity, like the bulk 
thermal conductivity, varies linearly with $T$, while if $\Gamma\ll 
T\ll\Delta_s$ (or if the impurity scatters are weak), the twin 
boundary thermal conductivity varies as $T^2$. The best way to 
identify twin boundary thermal conductivity experimentally would be on 
a highly twinned sample with all twin boundaries parallel to the (110) 
plane; measurements of thermal conductivities in the [110] and the 
$[1\bar{1}0]$ directions will then differ due to the thermal resistance 
of the twinning planes. Anisotropy in the basal plane thermal 
conductivity, which develops only below $T_c$, would be a clear 
indication of the particle-hole transformation in heat flow across 
twin boundaries.

\acknowledgments

This work was supported by the National Science and Engineering 
Research Council of Canada.

\begin{figure}
\caption{Twin boundary and nearest-neighbor hopping and pairing
amplitudes; $1\equiv(t_1,\Delta_1/2)$, 
$\bar{1}\equiv(t_1,-\Delta_1/2)$, $1'\equiv(t_1,\Delta'_1/2)$, 
$\bar{1'}\equiv(t_1,-\Delta'_1/2)$, $2\equiv(t_2,-\Delta_2/2)$ etc.}
\label{TB}
\end{figure}

\begin{figure}
\caption{Twin boundary and orthorhombic Fermi surfaces in the two 
twins. The capital letters A, B, ... denote the positions of nodes of 
the $d\pm s$ gaps. The sign of $\Delta({\bf k})$ in the different 
regions on  the Fermi surfaces is indicated as $+$ or $-$. } 
\label{FS}
\end{figure}

\end{document}